\documentclass[prb,twocolumn,aps,showpacs,superscriptaddress]{revtex4}
\usepackage{amsfonts}
\usepackage{graphicx}
\usepackage{textcomp}
\usepackage{txfonts}
\usepackage{bm}
\usepackage{CJK}

\begin{document}
\begin{CJK*}{GBK}{kai}

\title{Theory of magnetoelectric photocurrent generated by direct interband transitions in a semiconductor quantum well}

\author{Hai-Zhou Lu}\email{luhz@hku.hk}
\affiliation{Department of Physics and Center for Theoretical and Computational Physics, The University of Hong Kong, Pokfulam Road,
Hong Kong, China}

\author{Bin Zhou}
\affiliation{Department of Physics, Hubei University, Wuhan 430062, China}

\author{Fu-Chun Zhang}
\affiliation{Department of Physics and Center for Theoretical and Computational Physics, The University of Hong Kong, Pokfulam Road,
Hong Kong, China}

\author{Shun-Qing Shen}\email{sshen@hkucc.hku.hk}
\affiliation{Department of Physics and Center for Theoretical and Computational Physics, The University of Hong Kong, Pokfulam Road,
Hong Kong, China}

\date{\today }

\begin{abstract}
A linearly polarized light normally incident on a semiconductor quantum well with spin-orbit coupling may generate pure spin current via direct interband optical transition. An electric photocurrent can be extracted from the pure spin current when an in-plane magnetic field is applied, which has been recently observed in the InGaAs/InAlAs quantum well [Dai \emph{et al}. Phys. Rev. Lett. \textbf{104}, 246601 (2010)]. Here we present a theoretical study of this magnetoelectric photocurrent effect associated with the interband transition. By employing the density matrix formalism, we show that the photoexcited carrier density has an anisotropic distribution in $\mathbf{k}$ space, strongly dependent on the orientation of the electron wave vector and the polarization of the light. This anisotropy provides an intuitive picture of the observed dependences of the photocurrent on the magnetic field and the polarization of the light. We also show that the ratio of the pure spin photocurrent to the magnetoelectric photocurrent is approximately equal to the ratio of the kinetic energy to the Zeeman energy, which enables us to estimate the magnitude of the pure spin photocurrent. The photocurrent density calculated with the help of an anisotropic Rashba model and the Kohn-Luttinger model can produce all three terms in the fitting formula for measured current, with comparable order of magnitude, but discrepancies are still present and further investigation is needed.
\end{abstract}

\pacs{42.65.-k, 72.25.Dc, 72.40.+w, 73.63.Hs}

\maketitle

\section{Introduction}

Optical injection and detection of spin current have attracted a lot of interest recently in the context of non-magnetic semiconductor spintronics.\cite{Awschalom2007.natphys.3.153,Awschalom2009.physics.2.50}
Exciting progress has been made in recent years, such as the injection of spin-polarized currents by using the spin galvanic effect\cite{Ganichev2002.nature.417.153} and the circular photogalvanic effect,\cite{Ganichev2003.jpcm.15.R935} the observation of spin accumulations induced by the spin Hall effect with the help of the Kerr rotation\cite{Kato2004.science.306.1910} and the \emph{p}-\emph{n} junction light-emitting diode,\cite{Wunderlich2005.prl.94.047204} the inverse spin Hall effect,\cite{He2008.prl.101.147402,Wunderlich2009.natphys.5.675} and the detection of pure spin photocurrent by the second-harmonic effect.\cite{Werake2010,Wang2010.prl.104.256601}

Generation and detection of the pure spin current where there is no net charge current is one of the challenging issues in spintronics.
Optically, the generation of pure spin photocurrent has been realized by injecting orthogonally polarized one- and two-photon linear lights,\cite{Hache1997.prl.78.306,Bhat2000.prl.85.5432,Stevens2003.prl.90.136603,Hubner2003.prl.90.216601,Zhao2006.prl.96.246601}
or by the incidence of unpolarized or linearly polarized lights into bulk III-V semiconductors and quantum wells with spin-orbit couplings.\cite{Bhat2005.prl.94.096603,Tarasenko2005.jetp.81.231,Zhao2005.prb.72.201302R,Cui2007.apl.90.242115,Li2006.apl.88.162105,Zhou2007.prb.75.045339}
In the latter case, a linearly polarized light can be decomposed into circularly polarized lights with opposite helicities.
Due to the conservation of angular momentum, the circularly polarized lights with opposite helicities excite carriers with opposite spins, which are locked to opposite momenta owing to the spin-orbit coupling. As a result, the photo-injected carriers with opposite spins always have opposite velocities, leading to the pure spin photocurrent. One way to observed this underlying pure spin photocurrent is to apply an in-plane magnetic field, which will create an imbalance of spins. The spin imbalance leads to the asymmetry of electron velocities by the spin-orbit coupling and can extract an electric photocurrent from the pure spin photocurrent. The field-induced conversion from pure spin photocurrent to electric photocurrent (also referred to as the magneto-gyrotropic photogalvanic effect, MPGE) was systematically studied via the intra-band\cite{Ganichev2006.natphys.2.609,Belkov2005.jpcm.17.3405} and inter-subband\cite{Diehl2007.jpcm.19.436232} electron heating by THz and microwave radiations, in which spin-dependent scatterings are necessary for the observed charge current.

Very recently, the magnetoelectric photocurrent has been observed via direct interband optical transitions.\cite{Dai2010.prl.104.246601} Unlike the intra-band transition, in the direct interband transition the electron momentum is conserved, and the magnetoelectric photocurrent is a few orders of magnitude larger than that via the intraband transition. Therefore, it appears to be promising as a simple and practical method to generate and detect the spin current. Although symmetry analysis\cite{Belkov2005.jpcm.17.3405} can justify the observed electric photocurrent as a function of magnetic fields and polarization of the light, a microscopic description is still needed. In particular, the current induced by a parallel field is against intuition, considering the form of the spin-orbit coupling derived from the symmetry argument. Moreover, how to estimate the zero-field pure spin photocurrent from the magnetoelectric photocurrent is an interesting issue.

In this work, we study theoretically the generation of photocurrent via direct interband transitions excited by shedding a linearly polarized light normally into a semiconductor quantum well.
We find that the $\mathbf{k}$-space anisotropy of the carrier density provides an intuitive microscopic picture for the magnetoelectric photocurrent and the zero-field pure spin photocurrent. The origin of the observed dependences of the photocurrent on the magnetic field and the polarization of the light can be illustrated within this picture. The photocurrent density calculated with the help of a minimal model with an anisotropic Rashba model and the Kohn-Luttinger model is comparable with the experiment. Moreover, we propose an approach to estimate the magnitude of the undetectable pure spin photocurrent, which is generated at zero magnetic field by the same linear light, from the observed magnetoelectric photocurrent. Part of the results was very briefly reported in Ref. \onlinecite{Dai2010.prl.104.246601}.

The paper is organized as follows. In Sec. \ref{sec:experiment}, we review key features of the experimentally observed magnetoelectric photocurrent as a function of magnetic fields and the polarization of the linear light, and present a symmetry argument.
In Sec. \ref{sec:hamilton}, a minimal model of the quantum well is presented, as well as the electric-radiation interaction that accounts for the interband optical transitions. In Sec. \ref{sec:anisotropy}, we introduce the anisotropic photoexcited carrier density in $\mathbf{k}$ space, with the help of a standard density matrix formalism. In Sec. \ref{sec:picture}, we illustrate the origin of the field and polarization dependences of the magnetoelectric photocurrent in terms of the anisotropic photoexcited carrier density. In Sec. \ref{sec:quantitative}, we compare the calculated magnetoelectric photocurrent densities with the experiment. In Sec. \ref{sec:PSC}, the estimate of the underlying zero-field pure spin photocurrent is discussed. Finally, a summary is given in Sec. \ref{sec:conclusion}.

\section{\label{sec:experiment}Experiment and symmetry analysis}

\begin{figure}[tbph]
\centering \includegraphics[width=0.3\textwidth]{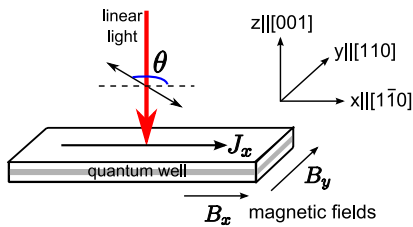}
\caption{Schematic illustration of the setup. A linearly polarized light (downward
arrow) is normally incident into a semiconductor quantum well grown along the [001] direction of zinc-blende materials. $\theta$ denotes the light polarization angle with respect to the $x$-axis. The $x$, $y$, and $z$ axes are defined along [1\={1}0], [110], and [001] crystallographic directions, respectively. $B_x$ ($B_y$) indicates the parallel (perpendicular) magnetic field with respect to the generated photocurrent $J_x$.}
\label{fig:setup}
\end{figure}

In this section, we will review the experiment on the magnetoelectric photocurrent generated via direct interband optical transitions by shedding a linearly polarized light normally into an InGaAs/InAlAs quantum well.\cite{Dai2010.prl.104.246601}
The experiment setup reported in Ref. \onlinecite{Dai2010.prl.104.246601} is schematically illustrated in Fig. \ref{fig:setup}.

In the experiment, the current density measured along the $x$ direction can be formulated as\cite{Dai2010.prl.104.246601}
\begin{equation}
j_{x}=c_{0}B_{y}+c_{y}B_{y}\cos 2\theta +c_{x}B_{x}\sin
2\theta ,  \label{eq:JxB}
\end{equation}%
where $B_x$ and $B_y$ are the magnetic fields along the $x$ and $y$ directions, respectively. $\theta$ is the polarization angle of the linearly polarized light with respect to the $x$ axis. $c_{0/x/y}$ are constant coefficients linearly scaled with the light power. The $x$ and $y$ axes here are defined along [1\={1}0] and [110] crystallographic directions of the zinc blende structure, respectively.

The experimental results are consistent with the symmetry argument of the $C_{\mathrm{2v}}$ point group.\cite{Belkov2005.jpcm.17.3405}
Inversion-asymmetric zinc blende heterostructures grown along [001] crystallographic direction usually have the $C_{\mathrm{2v}}$ point group symmetry. As a result, the electric photocurrent induced by in-plane magnetic fields can be phenomenologically written as (refer to Appendix \ref{sec:C2v} for details)
\begin{eqnarray}\label{eq:Jsymmetry}
j_{\mu }=\sum_{\nu }(\chi ^{\mu \mu \nu \bar{\nu}}B_{\mu }E_{\nu }E_{\overline{\nu }}
+\chi ^{\mu \bar{\mu}\nu \nu }B_{\bar{\mu}}E_{\nu }E_{\nu }),
\end{eqnarray}
where $\mu ,\nu $ run over $\{x,y\}$, and $\bar{\mu},\bar{\nu}=y$ if $\mu
,\nu =x$, and vice versa. $\chi $ is a fourth-rank pseudo tensor that
relates the electric photocurrent to the polarization electric field components
$(E_{x},E_{y})\propto (\cos \theta ,\sin \theta )$ of the incident light and the applied
magnetic fields $(B_{x},B_{y})$. One can readily check that Eq. (\ref{eq:Jsymmetry}) yields the same form as Eq. (\ref{eq:JxB}).

\begin{figure}[tbph]
\centering \includegraphics[width=0.4\textwidth]{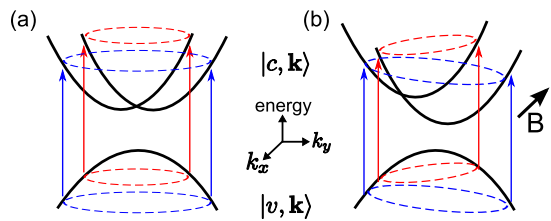}
\caption{(Color online) Schematic description of direct interband optical
transitions from a valence band $|v,\mathbf{k}\rangle$ to the spin-split
conduction bands $|c=\pm,\mathbf{k}\rangle$ at zero magnetic field (a)
and in the presence of a magnetic field $B$ (b). The states involved in the transitions form a set of constant energy contours
(denoted by dashed rings), which can be shifted along $k_{y}$ ($k_{x}$) in the presence of a magnetic field along $x$ ($y$) direction. }
\label{fig:schematic}
\end{figure}

Another consequence of the C$_{2v}$ symmetry is that when $x||$ [1\={1}0] and $y||$ [110], the invariants of the $C_{\mathrm{2v}}$ symmetry allow the spin-orbit coupling in linear $\mathbf{k}$ to be only in the form\cite{Ganichev2004.prl.92.256601,Giglberger2007.prb.75.035327} (refer to Appendix \ref{sec:C2v} for details)
\begin{eqnarray}\label{eq:soc}
H_{\mathrm{SOC}} = \lambda_y \sigma_x k_y - \lambda_x \sigma_y k_x ,
\end{eqnarray}
where $k_{x/y}$ is the wave vector along the $x$ ($y$) axis, the Pauli matrix $\sigma_{x/y}$ depicts the spin along the $x$ ($y$) direction, and $\lambda_{x/y}$ is the spin-orbit coupling coefficient along the $x$ ($y$) direction. Usually, the current along $x$ direction corresponds to the average shift of momenta of carriers along the $x$ direction. According to Eq. (\ref{eq:soc}), $k_x$ couples only to the spin along $y$ directions, thus it can only be shifted by a magnetic field along the $y$ direction. A magnetic field applied along $x$ direction can not shift the average $k_x$ [see Fig. \ref{fig:schematic}(b)]. According to this argument, the last term of Eq. (\ref{eq:JxB}) seems against intuition because it originates from a magnetic field along the $x$ direction.

The main task of this work is to present an intuitive picture to solve the above dilemma, as well as the magnetic field and polarization angle dependences of the electric photocurrent, in terms of the anisotropy of the photoexcited carrier density in $\mathbf{k}$ space. We will see that the three terms in Eq. (\ref{eq:JxB}) can be well interpreted by the microscopic picture presented in Fig. \ref{fig:rho0vx} (subfigures in boxes).

\section{\label{sec:hamilton}Model Hamiltonian}

In this work, we consider a minimal model that describes the energy bands near the band gap $\Delta$ of the quantum well. The advantage of the minimal model is that it is analytically solvable, thus physically transparent in delivering the physical picture. We expect that more sophisticated calculations, e.g. by using the 14-band $\mathbf{k}\cdot \mathbf{p}$ model\cite{Bhat2005.prl.94.096603} or the full band structure local density approximation,\cite{Nastos2007.prb.76.205113} can cover more details, but will not deviate from the physical picture obtained by the minimal model.

The total Hamiltonian is given by
\begin{eqnarray}\label{H_total}
H &=& H_C+H_V +\hat{V}(t).
\end{eqnarray}
where $H_C$ is for the conduction bands, $H_V$ for the valence bands, and $\hat{V}(t)$ for the electric-radiation that couples the conduction and valance bands by the linearly polarized light. The lowest conduction subbands are described by a two-dimensional free electron gas with the Rashba and Dresselhaus spin-orbit couplings and in-plane magnetic fields,
\begin{eqnarray}
H_{C} = \frac{\hbar^2}{2m^*}(k_x^2+k_y^2)+(\lambda_y\sigma_x
k_y-\lambda_x\sigma_y k_x )+ \mathbf{h}\cdot \boldsymbol{\sigma},
\end{eqnarray}
where $m^*$ is the effective mass of electron, $\hbar$ is Planck's constant over $2\pi$, $k_{x}$ ($k_y$) is the wave vector along the $x$ ($y$) axis, $\lambda_y=\alpha+\beta$, $\lambda_x=\alpha-\beta$, and $\alpha$ and $\beta$ are the Rashba and Dresselhaus spin-orbit coupling coefficients, respectively. Throughout the work, $x$ and $y$ axes are defined along [1\={1}0] and [110] crystallographic directions, respectively. $\boldsymbol{\sigma}=(\sigma_x,\sigma_y)$ is the vector of the Pauli matrices. $\mathbf{h}=(h_x,h_y)=\frac{1}{2}g_e\mu_B(B_x, B_y)$ is the Zeeman energies induced by the in-plane magnetic fields $(B_x,B_y)$, with $g_e$ the Land\'{e} \emph{g}-factor, and $\mu_B$ the Bohr magneton.
The eigenenergies and eigenstates of $H_C$ are given by
\begin{eqnarray}\label{cenergy}
\epsilon_{n\pm}(\mathbf{k}) = \frac{\hbar^2}{2m^*}k^2\pm\Lambda,
\end{eqnarray}
\begin{eqnarray}\label{cstates}
|+,\mathbf{k}\rangle = \left[
  \begin{array}{c}
    1/\sqrt{2} \\
    U \\
  \end{array}
\right],\
|-,\mathbf{k}\rangle =  \left[
  \begin{array}{c}
    -U^*  \\
   1/\sqrt{2} \\
  \end{array}
\right],
\end{eqnarray}
where $U =[\lambda_y k_y +h_x-i(\lambda_x
k_x-h_y)]/\sqrt{2}\Lambda $ and $\Lambda = \sqrt{(\lambda_y k_y+h_x)^2+(\lambda_x  k_x-h_y )^2}$.

The valence bands in Eq. (\ref{H_total}) are described by the isotropic ($\gamma_2=\gamma_3$) Kohn-Luttinger model,\cite{Peter}
\begin{eqnarray}\label{luttinger}
H_V &=&
-\frac{\hbar^2}{2m_e}[(\gamma_1+\frac{5}{2}\gamma_2)k^2-2\gamma_2(\mathbf{k}\cdot
\mathbf{S})^2],
\end{eqnarray}
where $m_e$ is the electron mass, $\mathbf{S}$ represents the $3/2$ spin operator matrices, $\gamma_1$ and $\gamma_2$ are two model parameters, and $\mathbf{k}=(k_x,k_y,k_z)$ is the wave vector. For simplicity, we approximate $\langle k_z\rangle= 0 $ and $\langle k_z^2\rangle \simeq (\frac{\pi}{d})^2$ in $H_V$ when considering the quantization along the $z$ direction of the quantum well, where $d$ is the thickness of the quantum well. The eigenenergies and states of the valence bands are given by
\begin{eqnarray}\label{venergy}
\epsilon_{1/2}(\mathbf{k})=\epsilon_{4/3}(\mathbf{k})=-\Delta-\frac{\hbar^2\gamma_1}{2m}(k^2+\langle
k_z^2\rangle)\pm \frac{\hbar^2\gamma_2}{2m} \Omega,
\end{eqnarray}
\begin{eqnarray}\label{vstates}
|1,\mathbf{k}\rangle &=& \left(
\begin{array}{c}
Q \\
0\\
K \\
0\\
\end{array}
\right),\
|2,\mathbf{k}\rangle =\left(
\begin{array}{c}
0\\
Q \\
0\\
-K \\
\end{array}
\right),
\nonumber\\
|3,\mathbf{k}\rangle &=& \left(
\begin{array}{c}
-K^* \\
0\\
Q \\
0\\
\end{array}
\right),\
|4,\mathbf{k}\rangle =\left(
\begin{array}{c}
0\\
K^* \\
0\\
Q \\
\end{array}
\right),\
\end{eqnarray}
where $\Delta$ is the energy gap, and
$Q = C_v\left(-k^2+2\langle k_z^2\rangle+\Omega\right)$, $K= \sqrt{3}C_vk_+^2$, $\Omega= 2\sqrt{k^4+\langle
k_z^2\rangle^2-k^2\langle k_z^2\rangle}$, and $C_v =\left[3k^4 +\left(k^2-2\langle
k_z^2\rangle-\Omega \right)^2\right]^{-1/2}$.

The electric-radiation interaction that couples the conduction and valence bands is given by,\cite{Peter}
\begin{eqnarray}
\hat{V}(t)=\frac{e}{m_e}\mathbf{\tilde{A}}\cdot\hat{\mathbf{p}},
\end{eqnarray}
where $-e$ is the electron charge and $\hat{\mathbf{p}}$ the electron momentum operator. Under the electric dipole approximation, the vector potential $\mathbf{\tilde{A}}$ is related to the electric field of the single-frequency light by
\begin{eqnarray}\label{A_eda}
\mathbf{\tilde{A}}(t) &=& \frac{1}{i\omega} [ \mathbf{E}(\omega) e^{-i\omega t}- \mathbf{E}^*(\omega) e^{i\omega t}]
\end{eqnarray}
where we consider a linearly polarized single-color light incident normally to the $x-y$ plane, with the frequency $\omega$ and the polarization electric vector
\begin{eqnarray}
\mathbf{E} &=& E_0(\cos\theta, \sin\theta, 0),
\end{eqnarray}
where $E_0$ is the amplitude and $\theta$ is the polarization angle with respect to the $x$ axis. With the help of the Poynting vector, the electric component $E_0$ can be evaluated from the energy flux of the laser by $\xi I  =(1/2)\sqrt{\varepsilon_0/\mu_0}E_0^2$,
where $\xi$ is the absorption efficiency and $I$ is the light intensity. $\varepsilon_0$ and $\mu_0$ are the dielectric constant and magnetic permeability in the vacuum, respectively. We always assume $\omega>\Delta$, therefore the dominant optical absorption mechanism is the direct interband transition. ``Direct" means that the wave vector $\mathbf{k}$ of the electron keeps unchanged in the transition.

We neglect the Zeeman effect in the valence bands, because the contribution to currents from holes is expected to be much smaller than electrons considering the short charge and spin lifetimes of holes in $n$-type quantum wells. Besides the lowest subbands $|+,\mathbf{k}\rangle$ and $|-,\mathbf{k}\rangle$, another pairs of conduction subbands $|+',\mathbf{k}\rangle$ and $|-',\mathbf{k}\rangle$ above them are also considered in the numerical calculation. Approximated as the energy levels of an infinitely depth potential well, $|\pm'\rangle $ are about $ 3(\hbar^2/2m^*)(\pi/d)^2$ above $|\pm\rangle $.
The differences in the effective mass and spin-orbit couplings between $|\pm\rangle $ and $|\pm'\rangle $ are neglected for simplicity. Due to the parity in the $z$ direction, conduction bands $|+,\mathbf{k}\rangle$ and $|-,\mathbf{k}\rangle$ couple to valence bands $|1,\mathbf{k}\rangle$ and $|4,\mathbf{k}\rangle$, while $|+',\mathbf{k}\rangle$ and $|-',\mathbf{k}\rangle$ to $|2,\mathbf{k}\rangle$ and $|3,\mathbf{k}\rangle$. For simplicity, we also neglect the diamagnetic contribution,\cite{Stern1967.pr.163.816,Gorbatsevich1993.JETP.57.580,Tarasenko2008} which should give qualitatively similar results to the Zeeman effect.

\section{\label{sec:anisotropy}Anisotropy of photoexcited carrier density in $\mathbf{k}$ space}

The photoexcited carrier density and all the physical quantities can be found within the density matrix formalism.
In this work, we will consider only the steady-state nonequilibrium photoexcited carrier density, which can be found by the approach similar to that of the second order nonlinear optical susceptibilities.\cite{Boyd}
We start with the Liouville equation of the density matrix,
\begin{eqnarray}\label{rho1}
\partial_t \rho_{nm} =
-\frac{i}{\hbar}\left[H,\hat{\rho}\right]_{nm}-\gamma\left(\rho_{nm}-\rho^{(0)}_{nm}\right),
\end{eqnarray}
where $n$ and $m$ run over the states in Eqs. (\ref{cstates}) and (\ref{vstates}), and $\gamma$ is a phenomenological damping parameter. $\rho^{(0)}_{nm}$ is the equilibrium density matrix before the light excitation. Because we assume a $n$-type quantum well, the valence bands are fully occupied $\rho^{(0)}_{v,v'}=\delta_{v,v'}$, where $v,v' \in \{1,2,3,4\}$. While the initial equilibrium density matrix of the conduction bands are described by the Fermi function,
\begin{eqnarray}
\rho^{(0)}_{c,c'}(\mathbf{k}) &=& f[\epsilon_c(\mathbf{k})]\delta_{c,c'}\equiv\frac{\delta_{c,c'}}{e^{[\epsilon_c(\mathbf{k})-E_{\mathrm{F}}]/k_{\mathrm{B}}T}+1},
\end{eqnarray}
where $c,c'\in \{+,-,+',-'\}$, $E_{\mathrm{F}}$ is the Fermi energy, $k_{\mathrm{B}}$ is the Boltzmann constant, and $T$ is the temperature. By treating $|c,\mathbf{k}\rangle$ and $|v,\mathbf{k}\rangle$ as unperturbed part and $\hat{V}(t)$ as perturbation, the perturbation equations up to the second order are given by
\begin{eqnarray}\label{rho012}
\partial_t \rho_{nm}^{(0)}&=&
-i\omega_{nm}\rho_{nm}^{(0)},\nonumber\\
\partial_t \rho_{nm}^{(1)}&=&
-i\omega_{nm}\rho_{nm}^{(1)}-\frac{i}{\hbar}\left[\hat{V},\hat{\rho}^{(0)}\right]_{nm}-\gamma\rho_{nm}^{(1)},\nonumber\\
\partial_t \rho_{nm}^{(2)}&=&
-i\omega_{nm}\rho_{nm}^{(2)}-\frac{i}{\hbar}\left[\hat{V},\hat{\rho}^{(1)}\right]_{nm}-\gamma\rho_{nm}^{(2)},
\end{eqnarray}
where $\omega_{nm}(\mathbf{k})=[\epsilon_n(\mathbf{k})-\epsilon_m(\mathbf{k})]/\hbar$. $\rho_{nm}$ are functions of $\mathbf{k}$ because the light is momentum-free under the electric dipole approximation. After a straightforward derivation, the leading order of the light induced steady-state density matrix for the conduction bands is found to be of the second order:
\begin{eqnarray}\label{rho(2)1}
\rho^{(2)}_{c,c'}(\mathbf{k})&=& \frac{\tau_e \pi e^2 }{\hbar^2\omega^2} \sum_{v}
[\mathbf{v}_{cv}(\mathbf{k})\cdot\mathbf{E}(\omega)][\mathbf{v}_{v c'}(\mathbf{k})\cdot\mathbf{E}^*(\omega)]\nonumber\\
&\times&\left[(1-f_{c})\delta(\omega-\omega_{cv})+(1-f_{c'})\delta(\omega-\omega_{c'v})\right],\nonumber\\
\end{eqnarray}
where the steady state is approximated by introducing the momentum relaxation time $\tau_e$,
\begin{eqnarray}
\rho^{(2)}_{c,c'}(\mathbf{k}) \approx \tau_e \partial_t \rho^{(2)}_{c,c'}(\mathbf{k},t).
\end{eqnarray}
$\tau_e=\mu_m m^*/e$ can be estimated from the mobility $\mu_m$ and the effective mass $m^*$.
Equation (\ref{rho(2)1}) recovers the result by employing the semiconductor optical Bloch equations\cite{Bhat2005.prl.94.096603} and Fermi's golden rules.\cite{Li2006.apl.88.162105} In the following, we will suppress the superscript of $\rho^{(2)}$ for simplicity.
By substituting the velocity $\mathbf{v}_{cv}$ in Eq. (\ref{rho(2)1}) by the position $\mathbf{r}_{cv}$ according to
\begin{eqnarray}\label{v and r}
\langle c,\mathbf{k}|\hat{v}_i |v,\mathbf{k}\rangle = \langle
c,\mathbf{k}|\frac{1}{i\hbar} [\hat{r}_i,H_0] |v,\mathbf{k}\rangle=i\omega_{cv}\langle c,\mathbf{k}|\hat{r}_i |v,\mathbf{k}\rangle,
\end{eqnarray}
Eq. (\ref{rho(2)1}) can also be expressed as
\begin{eqnarray}\label{rho(2)2}
&&\rho_{c,c'}(\mathbf{k})=\frac{\tau_e \pi e^2 }{\hbar^2\omega}\sum_{v}
[\omega_{c'v}(1-f_{c})\delta(\omega-\omega_{cv})\nonumber\\
&&+\omega_{cv}(1-f_{c'})\delta(\omega-\omega_{c'v})] [\mathbf{r}_{cv}(\mathbf{k})\cdot\mathbf{E}(\omega)][\mathbf{r}_{v c'}(\mathbf{k})\cdot\mathbf{E}^*(\omega)].\nonumber\\
\end{eqnarray}
By using the eigenstates in Eqs. (\ref{cstates}) and (\ref{vstates}) and considering the spatial wavefunctions of the eigenstates, the elements of the transition matrix $\mathbf{r}_{cv}\equiv(x_{cv}, y_{cv})=(x_{vc}^{\dag}, y_{vc}^{\dag})$ are found as
\begin{eqnarray}\label{eq:xvc}
&& x_{vc}\equiv\langle v,\mathbf{k}|\hat{x}|c,\mathbf{k}\rangle =a_{cv} \left[\begin{array}{cccc}
 \frac{1}{\sqrt{2}}(\frac{K^*}{\sqrt{3}}-Q)& (Q-\frac{K^*}{\sqrt{3}})U^*\\
 -(\frac{Q}{\sqrt{3}}+K^*)U & -\frac{1}{\sqrt{2}}(\frac{Q}{\sqrt{3}}+K^*)\\
 \frac{1}{\sqrt{2}}(K+\frac{Q}{\sqrt{3}})& -(K+\frac{Q}{\sqrt{3}})U^*\\
 (Q-\frac{K}{\sqrt{3}})U & \frac{1}{\sqrt{2}}(Q-\frac{K}{\sqrt{3}})\\
\end{array}\right]\nonumber\\
\end{eqnarray}
and
\begin{eqnarray}\label{eq:yvc}
&&y_{vc}\equiv \langle v,\mathbf{k}|\hat{y}|c,\mathbf{k}\rangle =i a_{cv}\left[
\begin{array}{cccc}
 \frac{1}{\sqrt{2}}(Q+\frac{K^*}{\sqrt{3}}) & -(Q+\frac{K^*}{\sqrt{3}})U^*\\
 (\frac{Q}{\sqrt{3}}-K^*)U &  \frac{1}{\sqrt{2}}(\frac{Q}{\sqrt{3}}-K^*) \\
 \frac{1}{\sqrt{2}}(\frac{Q}{\sqrt{3}}-K) & (K-\frac{Q}{\sqrt{3}} )U^*\\
 (\frac{K}{\sqrt{3}}+Q)U & \frac{1}{\sqrt{2}}(\frac{K}{\sqrt{3}}+Q)
\\\end{array}\right],\nonumber\\
\end{eqnarray}
where we have defined the effective dipole length
\begin{eqnarray}
a_{\mathrm{cv}} \equiv \langle 0,0|x|1,-1\rangle
\end{eqnarray}
with $|0,0\rangle$ and $|1,-1\rangle$ the spherical harmonic functions $Y_{0,0}$ and $Y_{1,-1}$, respectively.
$a_{cv}$ has the dimension of length.
$U$, $K$, and $Q$ in Eqs. (\ref{eq:xvc}) and (\ref{eq:yvc}) have been defined in Eqs. (\ref{cstates}) and (\ref{vstates}). Note that the optical selection rules owing to the $s-$ and $p$-wave natures of the conduction and valence bands have been incorporated in Eqs. (\ref{eq:xvc}) and (\ref{eq:yvc}). We neglect the density matrix of the valence bands because the charge and spin lifetimes of holes are much shorter than those of electrons for $n$-type quantum wells.

In this work, we will retain only the diagonal part of the density matrix because when ignoring the broadening of the light frequency $\omega$, the coherent contribution from the off-diagonal part of the density matrix can be neglected.\cite{Bhat2005.prl.94.096603}
The diagonal terms of the density matrix have clear physical meaning as the photoexcited carrier density. By substituting the polarization electric field vector $\mathbf{E}=E_0(\cos\theta, \sin\theta)$ into Eq. (\ref{rho(2)2}), the carrier density excited to conduction band $|c,\mathbf{k}\rangle$ can be written as the summation from different valence bands $|v,\mathbf{k}\rangle$
\begin{eqnarray}
\rho_{c,c}(\mathbf{k}) &=& \sum_{v}\rho_{cv,\mathbf{k}}
\end{eqnarray}
where $\rho_{cv,\mathbf{k}}$ represents the carrier density excited from valence band $|v,\mathbf{k}\rangle $ to conduction band $|c,\mathbf{k}\rangle$ as a function of the wave vector $\mathbf{k}$. $\rho_{cv,\mathbf{k}}$ can be divided into three terms according to their dependence on the polarization angle $\theta$,
\begin{eqnarray}\label{eq:rhock}
\rho_{cv,\mathbf{k}}&=& \rho_{cv,\mathbf{k}}^0+\rho_{cv,\mathbf{k}}^{\cos}\cos2\theta+\rho_{cv,\mathbf{k}}^{\sin}\sin2\theta,
\end{eqnarray}
where
\begin{eqnarray}
\rho_{cv,\mathbf{k}}^0&=&\xi I\frac{ 2\pi  \tau_e  e^2   }{\hbar^2}\sqrt{\frac{\mu_0}{\varepsilon_0}}
(1-f_{c})\delta(\omega-\omega_{cv})(|x_{cv}|^2+|y_{cv}|^2),\nonumber\\
\rho_{cv,\mathbf{k}}^{\cos}&=&\xi I\frac{2\pi  \tau_e  e^2   }{\hbar^2}\sqrt{\frac{\mu_0}{\varepsilon_0}}
(1-f_{c})\delta(\omega-\omega_{cv})(|x_{cv}|^2-|y_{cv}|^2),\nonumber \\
\rho_{cv,\mathbf{k}}^{\sin}&=&\xi I\frac{2\pi  \tau_e  e^2   }{\hbar^2}\sqrt{\frac{\mu_0}{\varepsilon_0}}
(1-f_{c})\delta(\omega-\omega_{cv})\mathrm{Re}(2x_{cv}y_{vc}).\nonumber\\
\end{eqnarray}
In the polar coordinates $(k_x,k_y)\equiv k(\cos\varphi, \sin\varphi)$, we can transform the delta function of $\omega$ into that of $k$,
\begin{eqnarray}\label{rho0deltak}
\rho_{cv,\mathbf{k}}^0&=&\xi I\frac{ 2\pi  \tau_e  e^2   }{\hbar^2}\sqrt{\frac{\mu_0}{\varepsilon_0}}
(1-f_{c})G[k_{cv}(\varphi)]\delta[k-k_{cv}(\varphi)]\nonumber\\
&&\times(|x_{cv}|^2+|y_{cv}|^2),\nonumber\\
\rho_{cv,\mathbf{k}}^{\cos}&=&\xi I\frac{2\pi  \tau_e  e^2   }{\hbar^2}\sqrt{\frac{\mu_0}{\varepsilon_0}}
(1-f_{c})G[k_{cv}(\varphi)]\delta[k-k_{cv}(\varphi)]\nonumber\\
&&\times(|x_{cv}|^2-|y_{cv}|^2),\nonumber \\
\rho_{cv,\mathbf{k}}^{\sin}&=&\xi I\frac{2\pi  \tau_e  e^2   }{\hbar^2}\sqrt{\frac{\mu_0}{\varepsilon_0}}
(1-f_{c})G[k_{cv}(\varphi)]\delta[k-k_{cv}(\varphi)]\nonumber\\
&&\times\mathrm{Re}(2x_{cv}y_{vc}),\nonumber\\
\end{eqnarray}
where we defined
\begin{eqnarray}
G[k_{cv}(\varphi)]\equiv \left.\frac{1}{\left|d\omega_{cv}(k,\varphi)/dk\right|} \right|_{k=k_{cv}(\varphi)} .
\end{eqnarray}
$G[k_{cv}(\varphi)]$ is of dimension of second$\cdot$meter$^{-1}$. $k_{cv}(\varphi)$ is the root of $\omega=\omega_{cv}(k,\varphi)$ for a given $\varphi$, i.e., the wave vectors on the constant energy contour (see Fig. \ref{fig:schematic}). The delta function $\delta[k-k_{cv}(\varphi)]$ restrict the values of $k$ on the constant energy contours.

\begin{figure}[tbph]
\centering \includegraphics[width=0.5\textwidth]{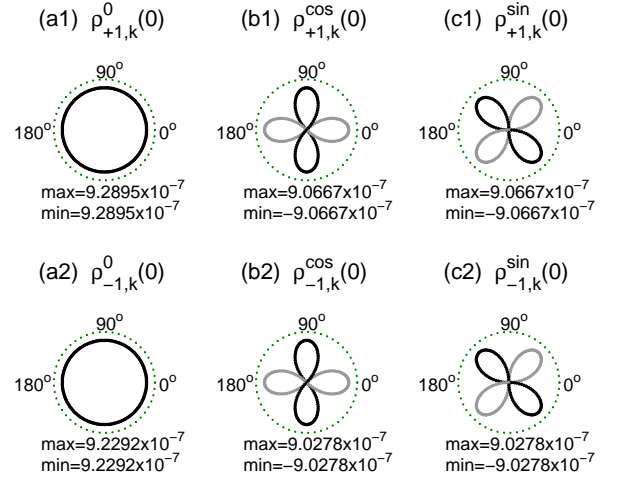}
\caption{At zero magnetic field ($B_x=0,B_y=0$), the calculated photoexcited carrier density $\rho_{cv,\mathbf{k}}$
in Eq. (\ref{rho0deltak}) as functions of the wave vector angle $\varphi$. Dark (light) represents positive (negative) values.
``max" (``min") indicates the maximum (minimum) values.  The $\rho_{cv,\mathbf{k}}^{0}$
term is always positive and overwhelms $\rho_{cv,\mathbf{k}}^{\cos/\sin}$. $\rho^{0,\cos,\sin}_{+1,\mathbf{k}}$ are in units of $(2\pi\xi I \tau_e  e^2a_{cv}^2\sqrt{\varepsilon_0/\mu_0}/\hbar^2)\times(\mathrm{second}/\mathrm{meter}$), and have a dimension of $\mathrm{meter}^{-1}$. Parameters: $T=$77 K, $E_{\mathrm{F}}=0.01$ eV, $\omega=0.8$ eV, $m^*=0.04m_e$, $\gamma_1=11.97$, $\gamma_2=4.36$, $\Delta=0.764$ eV,
$d=40$ nm, $\alpha=4.31$ meV$\cdot$nm, $\beta=0$. Numerical scheme is given in Appendix \ref{sec:Numerical}.}
\label{fig:rho0}
\end{figure}

In Fig. \ref{fig:rho0}, we show the calculated zero-field $\rho^0_{cv,\mathbf{k}}$, $\rho^{\cos}_{cv,\mathbf{k}}$, and $\rho^{\sin}_{cv,\mathbf{k}}$ as functions of $\varphi$, for the carriers excited from $|1,\mathbf{k}\rangle$ to $|\pm,\mathbf{k}\rangle$. It shows that $\rho^0_{cv,\mathbf{k}}$ is an isotropic function of $\varphi$, while $\rho^{\cos}_{cv,\mathbf{k}}$ and $\rho^{\sin}_{cv,\mathbf{k}}$ depend on $\cos2\varphi$ and $\sin2\varphi$, respectively, indicating the total carrier density excited by the linear light is anisotropic in $\mathbf{k}$ space. $\rho^0_{cv,\mathbf{k}}$ are always positive and overwhelm $\rho^{\cos,\sin}_{cv,\mathbf{k}}$ in magnitude, so the total photoexcited carrier density is physically positive. In addition, as indicated by the ``max"s and ``min"s, the carriers excited from the same valence band have different carrier density on the spin-split conduction bands $|+,\mathbf{k}\rangle$ [Figs. \ref{fig:rho0}(a1)-(c1)] and $|-,\mathbf{k}\rangle$ [Figs. \ref{fig:rho0}(a2)-(c2)]. This difference is due to the splitting of the constant energy contours for $|+,\mathbf{k}\rangle $ and $|-,\mathbf{k}\rangle $ by the spin-orbit coupling, as shown in Fig. \ref{fig:schematic}(a). Fig. \ref{fig:rho0} shows only the photoexcited carrier density from $|1,\mathbf{k}\rangle$ to $|\pm,\mathbf{k}\rangle$. Other pairs of conduction and valence bands also have similar anisotropic photoexcited carrier density in $\mathbf{k}$ space.

Summarizing the above, the carrier density excited via direct interband transitions from valence band $|v,\mathbf{k}\rangle$ to conduction band $|c,\mathbf{k}\rangle$ by the normal incidence of a linearly polarized light can be expressed as
\begin{eqnarray}\label{eq:rho0varphi}
\rho_{cv,\mathbf{k}}= \rho_{cv,\varphi}^0+\rho_{cv,\varphi}^{\cos}\cos2\varphi\cos2\theta+\rho_{cv,\varphi}^{\sin} \sin2\varphi\sin2\theta,
\end{eqnarray}
where $\rho_{cv,\varphi}^{0,\cos,\sin}$ in general are functions of $\varphi$.
In a weak in-plane magnetic field (say, less than 1 tesla), $\rho_{cv,\varphi}^{0,\cos,\sin}$ slightly depend on $\varphi$.
In the absence of the magnetic field and the anisotropy of the spin-orbit couplings,
$\rho_{cv,\varphi}^{0,\cos,\sin}$ become independent on $\varphi$.
In this case, Eq. (\ref{eq:rho0varphi}) can be written as
\begin{eqnarray}\label{eq:rho0}
\rho_{cv,\mathbf{k}}= \rho_{cv}^0+\rho_{cv}^{\cos}\cos2\varphi\cos2\theta+\rho_{cv}^{\sin}\sin2\varphi\sin2\theta,
\end{eqnarray}
where $\rho_{cv}^{0,\cos,\sin}$ are constants of dimension meter$^{-1}$.

The anisotropy of the photoexcited carrier density is the core of this paper. It can naturally account for the field and polarization dependence of the magnetoelectric photocurrent\cite{Dai2010.prl.104.246601} and the pure spin photocurrent.\cite{Li2006.apl.88.162105,Zhou2007.prb.75.045339}

\section{\label{sec:IB}Magnetic-field induced electric photocurrent}

\begin{widetext}

\begin{figure}[tbph]
\centering \includegraphics[width=0.7\textwidth]{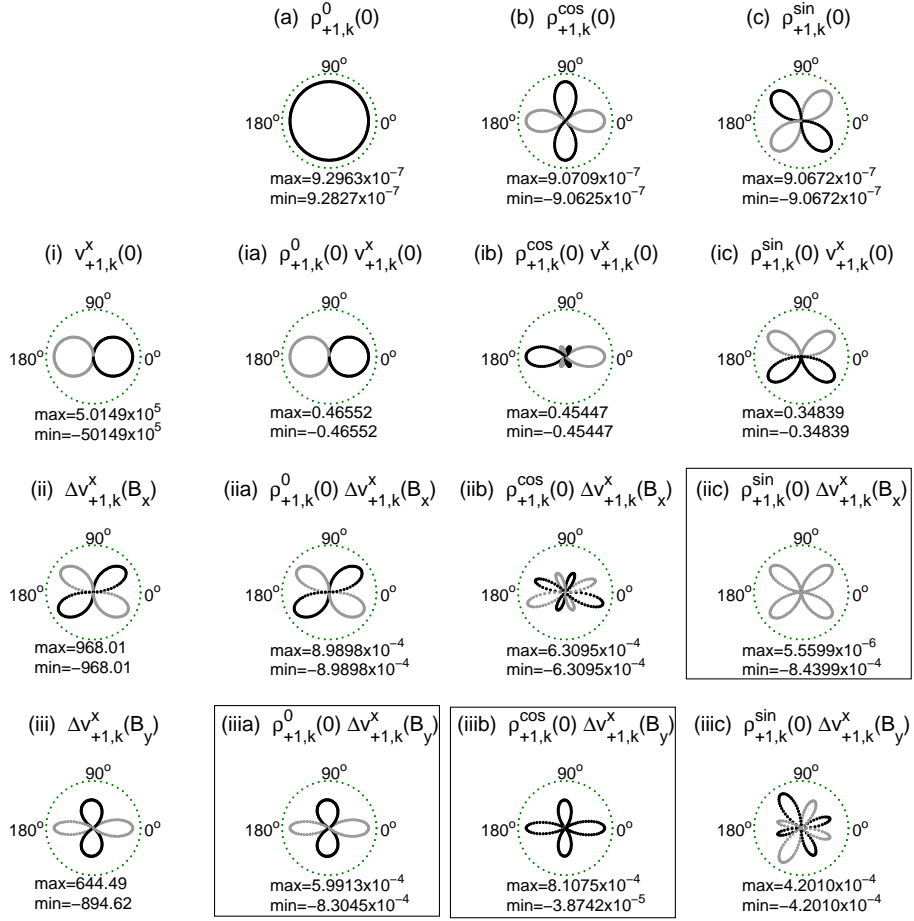}
\caption{The product table of the photoexcited carrier density and velocity as functions of wave vector angle $\varphi$. Only the carriers excited from $|1,\mathbf{k}\rangle$ to $|+,\mathbf{k}\rangle$ are shown. [(a)-(c)] The zero-field photoexcited carrier density $\rho_{+1,\mathbf{k}}^{0,\cos,\sin}$. [(i)-(iii)] The zero-field velocity (i), and its variation with $B_x$ (ii) and $B_y$ (iii), respectively. ``$(0)$" stands for the values without the magnetic field, and ``$\Delta...(B_{\nu}) $" for the variation upon applying the field $B_{\nu}$. Dark (light) represents that the value is positive (negative). ``max" (``min") indicates the maximum (minimum) values. The boxes mark the origins of the three terms in Eq. (\ref{eq:JxB}). The parameters are given in Sec. \ref{sec:quantitative}. $\beta=1$ meV$\cdot$nm is assumed. Numerical scheme is given in Appendix \ref{sec:Numerical}.}
\label{fig:rho0vx}
\end{figure}

\end{widetext}

\subsection{\label{sec:picture}Origin of $c_0$, $c_x$, and $c_y$}

The electric photocurrent density along $\mu(\in\{x,y\})$ axis can be found by summing the velocities of the photoexcited carriers
\begin{eqnarray}\label{jmusum}
j_{\mu} &=& -e\sum_{c,v,\mathbf{k}}\rho_{cv,\mathbf{k}} v^{\mu}_{cv,\mathbf{k}}.
\end{eqnarray}
From the eigenenergies Eq. (\ref{cenergy}), the velocity along the $x$ direction for conduction bands $|\pm,\mathbf{k}\rangle $ can be found as
\begin{eqnarray}
v^x_{\pm\mathbf{k}} \equiv \frac{1}{\hbar}\frac{\partial \epsilon_{\pm}}{\partial k_x} = \frac{\hbar}{m^*}k_x\pm \frac{\alpha}{\hbar}\frac{(\alpha k_x -h_y)}{\sqrt{(\alpha k_y+h_x)^2+(\alpha  k_x-h_y )^2}}.
\end{eqnarray}
Above and hereafter, we will ignore $\beta$ in the analytical expressions, and take it into account only in the numerical calculation.
By rewriting $k_x=k\cos\varphi$ and $k_y=k\sin\varphi$ in polar coordinates $(k,\varphi)$ and expanding the velocities up to the linear order in $h_x$ and $h_y$, we have
\begin{eqnarray}\label{velocitytaylor}
v_{\pm\mathbf{k}}^{x}&\simeq&(\frac{\hbar}{m^{\ast}}k\pm\frac{\alpha}{\hbar})\cos\varphi\mp\frac{\sin2\varphi}{2\hbar k}h_{x}\mp\frac{\sin^{2}\varphi}{\hbar k}h_{y}.
\end{eqnarray}
$v_{cv,\mathbf{k}}^{\mu}$ in the current density formula (\ref{jmusum}) is related to Eq. (\ref{velocitytaylor}) by restricting $k$ on the constant energy contours $k_{cv}$.

Rewriting the summation in Eq. (\ref{jmusum}) into an integral in polar coordinates, putting into Eqs. (\ref{eq:rho0varphi}) and (\ref{velocitytaylor}),  and performing the integral over $k$, we obtain an integral over $\varphi$
\begin{eqnarray}\label{jxpolar}
&&j_{x}= -\frac{e}{(2\pi)^2}\sum_{c ,v}\int_0^{2\pi}d\varphi k_{cv}(\varphi) \nonumber\\
&&\times[\rho_{cv,\varphi}^0 + \rho_{cv,\varphi}^{\cos}\cos2\varphi\cos2\theta + \rho_{cv,\varphi}^{\sin}\sin2\varphi\sin2\theta ]\nonumber\\
&& \times [(\frac{\hbar}{m^{\ast}}k_{cv}(\varphi)+c\frac{\alpha}{\hbar})\cos\varphi-c\frac{\sin2\varphi}{2\hbar k_{cv}(\varphi)}h_{x}-c\frac{\sin^{2}\varphi}{\hbar k_{cv}(\varphi)}h_{y}],\nonumber\\
\end{eqnarray}
where $\varphi \equiv \arctan (k_{y}/k_{x})$, $k^2\equiv k_x^2+k_y^2$, and $k_{cv}(\varphi)$ are the roots of $\omega=\omega_{cv}(k,\varphi)$. In general, $k_{cv}(\varphi)$ and $\rho^{0,\cos,\sin}_{cv,\varphi}$ are functions of $\varphi$ when $\mathbf{B}\neq 0$ and $\beta\neq 0$.
At this moment, if we ignore their $\varphi$ dependence, the above integral can be readily performed,
\begin{eqnarray}\label{jxpolarBxBy}
&&j_{x}(B_x,B_y) \simeq   h_{y} \frac{e}{(2\pi)^2}(\sum_{c ,v}\frac{c}{\hbar } \rho_{cv}^0) \int_0^{2\pi}d\varphi \sin^{2}\varphi
\nonumber\\
&&+ h_{y} \frac{e}{(2\pi)^2}(\sum_{c ,v}\frac{c}{\hbar }  \rho_{cv}^{\cos} ) \cos2\theta \int_0^{2\pi}d\varphi \cos2\varphi\sin^{2}\varphi\nonumber\\
&& + h_{x}\frac{e}{(2\pi)^2}(\sum_{c ,v}  \frac{c}{2\hbar } \rho_{cv}^{\sin})\sin2\theta
 \int_0^{2\pi}d\varphi\sin2\varphi\sin2\varphi , \nonumber\\
&& =  (\frac{e}{4\pi\hbar }\sum_{c ,v}c \rho_{cv}^0)h_{y}
- ( \frac{e}{8\pi\hbar }\sum_{c ,v}c\rho_{cv}^{\cos})h_{y}\cos2\theta \nonumber\\
&&+(\frac{e}{8\pi\hbar}\sum_{c ,v}  c \rho_{cv}^{\sin}) h_{x}\sin2\theta.
\end{eqnarray}
The resulting three nonzero terms immediately recovers the form of the experimental fitting formula (\ref{eq:JxB}). In this way, a clear relation between the anisotropy of the photoexcited carrier density and the magnetoelectric photocurrent density is established.

Above we ignore the $\varphi$ dependence of $\rho^{0,\cos,\sin}_{cv}$ and $k_{cv}$ resulting from finite $\mathbf{B}$ and $\beta$.
Their influences can be taken into account numerically. In Fig. \ref{fig:rho0vx} we present a product table of calculated $v^x_{+1,\mathbf{k}}$ and $\rho^{0,\cos,\sin}_{+1,\mathbf{k}}$ as functions of $\varphi$, for finite $\mathbf{B}$ and $\beta$ (refer to Appendix \ref{sec:Numerical} for the numerical scheme). The three nonzero contributions in Eq. (\ref{jxpolarBxBy}) are marked by the boxed sub-figures (iic), (iiia), and (iiib). Except the three nonzero terms, note that all the other sub-figures in Fig. \ref{fig:rho0vx} always have equal weight of positive and negative lobes. In other words, they yield zero when integrating over $\varphi$, and give no contribution to the electric current.
All the pairs of conduction and valence bands have the same behavior, and all of them add up (actually cancel with each other) to give the total net electric current.

\begin{figure}[tbph]
\centering \includegraphics[width=0.5\textwidth]{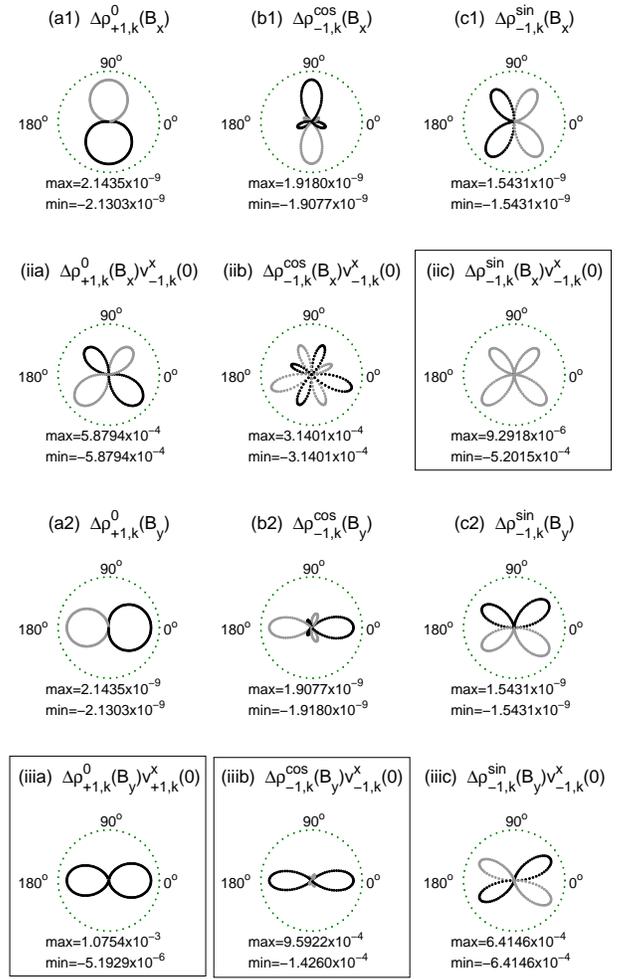}
\caption{The product table of the field-induced variation of the photoexcited carrier density and the zero-field velocity as functions of $\varphi$. The legend is the same as in Fig. \ref{fig:rho0vx}. The boxed three terms, when added to the boxed terms in Fig. \ref{fig:rho0vx}, gives the ``$2\eta$" factor in front of Eq. (\ref{jmubnuapprox}). All parameters are the same as those in Fig. \ref{fig:rho0vx}. Numerical scheme is given in Appendix \ref{sec:Numerical}.}
\label{fig:rhoBvx0}
\end{figure}

In the analytic result (\ref{jxpolarBxBy}), we have neglected the variation of $\rho_{cv,\mathbf{k}}^{0,\cos,\sin}$ when applying the magnetic field. In general, the charge current $j_{\mu}$ induced by $B_{\nu}$ can be expanded as
\begin{eqnarray}\label{eq:expansion}
j_{\mu}(B_{\nu})\simeq-e\sum_{c,v,\mathbf{k}}[\rho_{cv,\mathbf{k}}(0)v^{\mu}_{c\mathbf{k}}(0)+\rho_{cv,\mathbf{k}}(0)\Delta v^{\mu}_{c\mathbf{k}}(B_{\nu})\nonumber\\
+\Delta\rho_{cv,\mathbf{k}}(B_{\nu})v^{\mu}_{c\mathbf{k}}(0) +\Delta\rho_{cv,\mathbf{k}}(B_{\nu})\Delta v^{\mu}_{c\mathbf{k}}(B_{\nu})]+\mathcal{O}(B_{\nu}^2),
\end{eqnarray}
where ``$(0)$" stands for ``at zero magnetic field", and ``$\Delta...(B_{\nu}) $" for the variation when applying the magnetic field $B_{\nu}$ along the $\nu$ direction. We already illustrated that the first term vanishes, and the second term is consistent with the experiment.
Besides, one can expect that the third term, i.e., the magnetic field-induced variation of the photoexcited carrier density,
gives a contribution of the same order as the second term [see Fig. \ref{fig:rhoBvx0}], and the last term is ignorably small. Therefore, we approximate the photocurrent density induced by the magnetic field by
\begin{eqnarray}\label{jmubnuapprox}
j_{\mu}(B_{\nu}) &\simeq & (\eta \frac{e}{2\pi\hbar }\sum_{c ,v}c \rho_{cv}^0) h_{y}
- (\eta  \frac{e}{4\pi\hbar }\sum_{c ,v}c\rho_{cv}^{\cos})h_{y}\cos2\theta\nonumber\\
&&+ (\eta \frac{e}{4\pi\hbar}\sum_{c ,v}  c \rho_{cv}^{\sin})h_{x}\sin2\theta,
\end{eqnarray}
where an extra ``$2\eta$" has been multiplied to account for the summation of the second and third terms of Eq. (\ref{eq:expansion}).

\subsection{\label{sec:quantitative}Comparison with experiment}

Above we show that the photoexcited carrier density in Eq. (\ref{eq:rho0varphi}) explains the origin of $c_0$, $c_y$, and $c_x$ terms in the fitting formula (\ref{eq:JxB}) of the magnetoelectric photocurrent. Now we compute $c_{0/x/y}$, and see how close the minimal model can be when comparing with the experiment.

We choose a set of parameters close to the experiment. From the experiment, we have that the temperature $T=$77 K, the Fermi level $E_{\mathrm{F}}=0.01$ eV, the light frequency $\omega=0.8$ eV, the band gap $\Delta \approx 0.764$ eV, the quantum well thickness $d=40$ nm, and the Rashba spin-orbit coupling constant $\alpha=4.31$ meV$\cdot$nm.\cite{Dai2010.prl.104.246601} The Luttinger model parameters $\gamma_1=11.97$ and $\gamma_2=4.36$ are adopted from those for Ga$_{0.47}$In$_{0.53}$As,\cite{Winkler} which has the similar components as those in the experiment (Ga$_{1-x}$In$_{x}$As, $x=0.53\sim 0.59$ by graded doping). The momentum relaxation time can be estimated by $\tau_e=\frac{m^*\mu_m }{e}\approx 2$ ps, with the effective mass $m^*\approx 0.04$ and the mobility $\mu_m\approx 84000\ \mathrm{cm}^2/\mathrm{(V s)}$ at 77 K (about 7 times larger than that at room temperature \cite{Dai2010.prl.104.246601}).
The effective dipole length $a_{\mathrm{cv}}\approx 6.7\AA$, is approximated by that for GaAs.\cite{Wang2010.prl.104.256601}
Considering a reflectance of 0.3 and the absorption coefficients $9\times 10^{3} \mathrm{cm}^{-1}$, we obtained the absorption efficiency $\xi = (1-0.3)[1-\exp(-9\times 10^{3} \mathrm{cm}^{-1}\times 40 \mathrm{nm})] \approx 2.5 \% $. With the light power $P=15$ mW and the light spot radius $5\mu $m, the light intensity is found as $I\approx 1.91\times 10^{8}$ W/m$^2$. The experiment observed that $|c_x|\neq|c_y|$; this may be due to a finite Dresselhaus spin-orbit coefficient $\beta$, so that the spin-orbit couplings along the $x$ and $y$ directions are different, i.e., $\lambda_x\neq\lambda_y$, since $\lambda_x=\alpha-\beta$ and $\lambda_y=\alpha+\beta$. The value of $\beta$ is unknown to the experiment. In this following calculations, $\beta$ will be chosen as a variable parameter.

\begin{figure}[tbph]
\centering \includegraphics[width=0.5\textwidth]{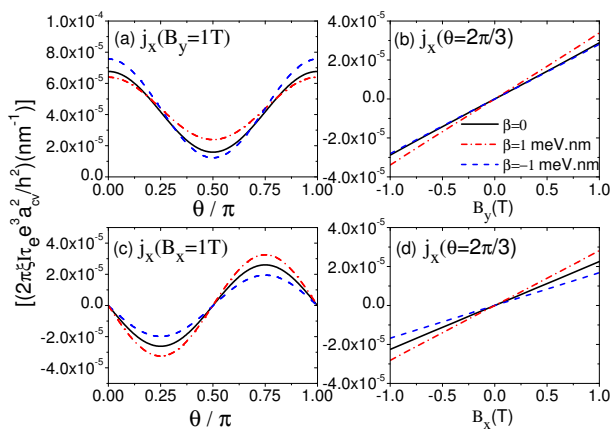}
\caption{(Color online) Calculated $j_x$ for different $\beta$. (a) $j_x(\theta)$ at $B_y=1$ T. (b) $j_x(B_y)$ at $\theta=120^{\circ}$. (c) $j_x(\theta)$ at $B_x=1$ T. (d) $j_x(B_x)$ at $\theta=120^{\circ}$. Parameters are given in Sec. \ref{sec:quantitative}. }
\label{fig:jx}
\end{figure}

In Fig. \ref{fig:jx}, we present the numerically calculated photocurrent densities $j_x$ as functions of $B_{x}$, $B_{y}$ and $\theta$. We also compare the results for positive, zero, and negative $\beta$. Qualitatively, the calculated results capture the main features of the experiment. Fig. \ref{fig:jx}(b) and (d) show that, the current densities are linearly scaled with the magnetic fields. At zero magnetic field, there is no current. Fig. \ref{fig:jx}  (a) and (c) show that, the parallel field leads to only $\sin2\theta$ dependence while the perpendicular field induces only the constant and $\cos2\theta$ dependences. In summary, the calculated current density can be formulated in the same form as Eq. (\ref{eq:JxB}).

\begin{table}[tbph]
\caption{Comparison between the theory and experiment for the parameters $c_{0}/I$, $c_{x}/I$, and $c_{y}/I$ in formula (\ref{eq:JxB}).
$I$ is the light intensity. $c_{0/x/y}/I$ are in units of $10^{-14} \times$ (A/m)/(T$\cdot$W/m$^2$). Parameters are given in Sec. \ref{sec:quantitative}.}
\label{tab:compare}
\begin{tabular}{l cccccccc}
  \hline
  \hline
  &    $c_0/I$  & $c_y/I$  & $c_x/I$ & $|c_y/c_x|$ \\
     \hline
 Experiment\cite{Dai2010.prl.104.246601} &   $10.2$ &   $-1.74$ &   $-0.94$ & 1.85 \\
 \hline
Theory ($\beta/\alpha=-1/3$)    &   $2.2$	 &   $1.6$	&  $-0.8$ & 2 \\
Theory ($\beta=0$) & 	$1.9$  &   $1.2$  &  $-1.2$  & 1\\
Theory ($\beta/\alpha=1/3$) & 	$2.2$  &  $0.8$	&  $-1.6$ & 0.5 \\
  \hline
  \hline
\end{tabular}
\end{table}

Now we make some quantitative comparisons. In Fig. \ref{fig:jx}, the current densities are in units of $\zeta\approx 0.09 $ A/m [see Eq. (\ref{j0units})]. Fig. \ref{fig:jx} shows that the dominant term is $j_x(B_y)$, consistent with the experiment. $j_x(B_y)$ can be as large as $ 8\times 10^{-5} \zeta $ when $B_y=1$ T, i.e., about $ 0.72 \times 10^{-5}$ A/m, comparable with the experimental estimate $\sim 2\times 10^{-5}$ A/m.\cite{Dai2010.prl.104.246601} The calculated current is smaller, probably because only limited bands are included in the calculation.
In table \ref{tab:compare}, we compare the calculated $c_0$, $c_y$, and $c_x$ with the experiment.
The calculated $c_y$ and $c_x$ are comparable with $c_0$, while in the experiment $c_0$ is almost an order of magnitude larger than $c_y$ and $c_x$. In Sec. \ref{sec:IB}, we have shown that $c_y$ and $c_x$ terms come from the photoexcited carrier $\rho_{cv}^{\cos,\sin}$, which originate from the quantum interference between two circular components of a linearly polarized light.\cite{Bhat2005.prl.94.096603} In contrast to the ideal situation assumed in the theory, the interference effect may be suppressed in the experiment, then $c_y$ and $c_x$ are reduced and $c_0$ is enhanced by a non-interference contribution.
Besides, the sign of the calculated $c_y$ is opposite to that of the experiment.
This may be attributed to the difference in band symmetries between the experiment and the model.
Considering the simplicity of the minimal model, we expect that more sophisticated models may cover more reliable details, but the minimal model is enough to offer a reasonable physical picture for the experiment.

\section{\label{sec:PSC}Zero-field pure spin photocurrent}

\subsection{Polarization dependence of zero-field Pure spin photocurrent}

\begin{widetext}

\begin{figure}[htbp]
\centering \includegraphics[width=0.7\textwidth]{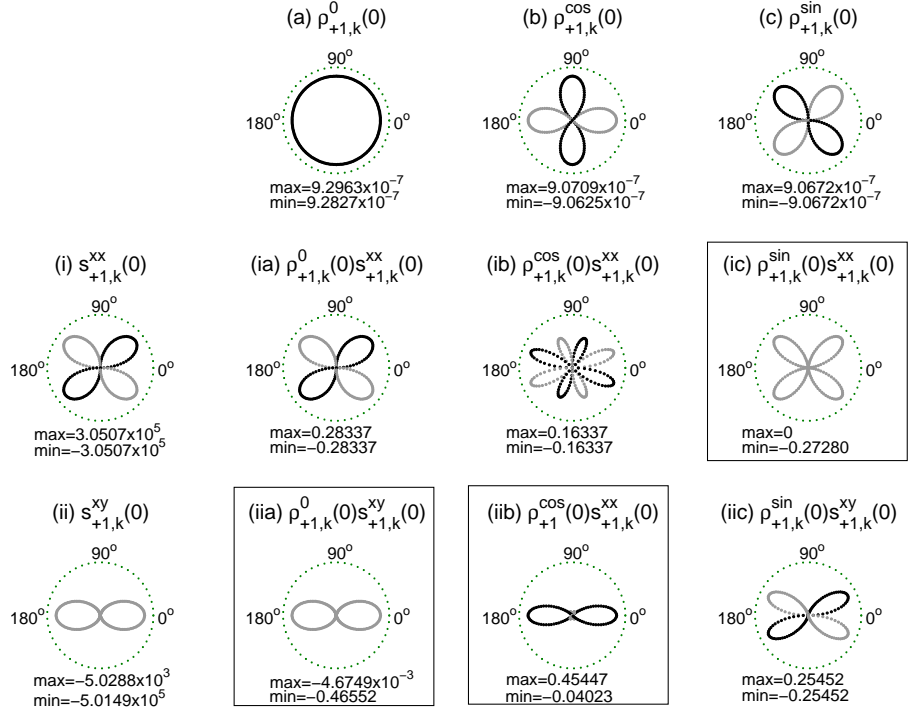}
\caption{The product table of the photoexcited carrier density and spin velocities as functions of wave vector angle $\varphi$ at zero magnetic field. Only the carriers excited from $|1,\mathbf{k}\rangle$ to $|+,\mathbf{k}\rangle$ are shown. [(a)-(c)] The photoexcited carrier density $\rho_{+1,\mathbf{k}}$. [(i)-(ii)] The spin velocities flowing along $x$ axis, while with the spin pointing along $x$ [(i)] and $y$ [(ii)] directions, respectively. Dark (light) represents that the value is positive (negative). ``max" (``min") indicates the maximum (minimum) values. The boxes mark the origins of the three terms in Eq. (\ref{eq:PSC}). The parameters are given in Sec. \ref{sec:quantitative}. $\beta=1$ meV$\cdot$nm is assumed. Numerical scheme is given in Appendix \ref{sec:Numerical}.}
\label{fig:rho0sxxsxy}
\end{figure}

\end{widetext}

Both the symmetry argument of the $C_{\mathrm{2v}}$ group\cite{Tarasenko2005.jetp.81.231} and theoretical calculations\cite{Bhat2005.prl.94.096603,Li2006.apl.88.162105,Zhou2007.prb.75.045339} have pointed out the generation of pure spin currents by the linearly polarized or unpolarized lights, and they can be expressed as functions of $\theta$:
\begin{eqnarray}\label{eq:PSC}
j_x^{y} = I_0 + I_1 \cos2\theta,\ \ j_{x}^{x} = I_2\sin2\theta,
\end{eqnarray}
where $I_{0,1,2}$ are constant coefficients. In this section, we will show how the anisotropy of the photoexcited carrier density is related to the $\theta$-dependence of the pure spin current.

The spin photocurrent density flowing along the $\mu$ ($\in\{x,y\}$) direction with spin polarized along the $\nu$ ($\in\{x,y\}$) direction can be found by
\begin{eqnarray}\label{jmunudefinition}
j_{\mu}^{\nu} =\frac{\hbar}{2}\sum_{c,v,\mathbf{k}}\rho_{cv,\mathbf{k}} s_{cv,\mathbf{k}}^{\mu\nu},
\end{eqnarray}
where we also retain only the diagonal density matrix, and $s_{cv,\mathbf{k}}^{\mu\nu}$ is the spin velocity for the carriers excited from valence band $|v,\mathbf{k}\rangle$ to conduction band $|c,\mathbf{k}\rangle$. The spin velocity operator is defined as
\begin{eqnarray}
\hat{s}^{\mu\nu} &=& \frac{1}{2}\{\sigma_{\nu}, \frac{1}{\hbar}\frac{\partial H_C}{\partial k_{\mu}}\},
\end{eqnarray}
where $\sigma_{\nu}$ is the Pauli matrix. The zero magnetic field expectation value of the spin velocity for conduction band $|c,\mathbf{k}\rangle$, defined as $s^{\mu\nu}_{\pm\mathbf{k}}\equiv\langle \pm,\mathbf{k}|\hat{s}^{\mu\nu}|\pm,\mathbf{k}\rangle$, can be found in polar coordinates as,
\begin{eqnarray}\label{smunu}
s^{xx}_{\pm\mathbf{k}} = \pm \frac{\hbar}{2m^*}k \sin2\varphi,\ s^{xy}_{\pm\mathbf{k}} =
\mp \frac{\hbar}{m^*}k \cos^2\varphi -\frac{\alpha}{\hbar},\nonumber\\
s^{yx}_{\pm\mathbf{k}} =
\pm \frac{\hbar}{m^*}k\sin^2\varphi+\frac{\alpha}{\hbar},\
s^{yy}_{\pm\mathbf{k}} = \mp \frac{\hbar}{2m^*} k \sin2\varphi.
\end{eqnarray}
$s_{cv,\mathbf{k}}^{\mu\nu}$ in the spin current density formula (\ref{jmunudefinition}) is related to Eq. (\ref{smunu}) by restricting $k$ on the constant energy contours between valence band $|v,\mathbf{k}\rangle$ and conduction band $|c,\mathbf{k}\rangle$.

Similar to the current density, the spin current density can also be rewritten into an angle integral,
\begin{eqnarray}\label{jmunuangleintegral}
j_{\mu}^{\nu}(0) = \frac{\hbar}{2}\frac{1}{(2\pi)^2} \sum_{c,v}  \int_0^{2\pi }d\varphi k_{cv}(\varphi)\ s_{cv}^{\mu\nu}(k_{cv},\varphi)\nonumber\\
\times \left[\rho_{cv,\varphi}^0 + \rho_{cv,\varphi}^{\cos}\cos2\varphi\cos2\theta + \rho_{cv,\varphi}^{\sin}\sin2\varphi\sin2\theta \right].
\end{eqnarray}
If we ignore the $\varphi$ dependence of $\rho_{cv,\varphi}^{0,\cos,\sin}$ and $k_{cv}(\varphi)$,
the above integral yields
\begin{eqnarray}\label{jxxresult}
j_{x}^{x}(0) &=& \frac{\hbar}{2}\frac{1}{4\pi}\sin2\theta \sum_{c,v} c  \frac{\hbar}{2m^*}k_{cv}^2 \rho_{cv}^{\sin} ,\nonumber\\
j_{x}^{y}(0) &=& -\frac{\hbar}{2}\frac{1}{2\pi}\sum_{c,v}
 \left[c \frac{\hbar}{2m^*}k_{cv}^2  \rho_{cv}^0 +\frac{\alpha}{\hbar} \rho_{cv}^0 k_{cv} \right]\nonumber\\
&& -\frac{\hbar}{2}\frac{1}{2\pi}\sum_{c,v}
c \frac{1}{2}\frac{\hbar}{2m^*}k_{cv}^2  \rho_{cv}^{\cos}\cos2\theta ,
\end{eqnarray}
which give the $\theta$ dependence in Eq. (\ref{eq:PSC}).
In the presence of a finite $\beta$, the result is not affected qualitatively,
as shown in Fig. \ref{fig:rho0sxxsxy}, where the nonzero contributions to Eq. (\ref{eq:PSC}) are marked by the boxes.

\subsection{Quick estimate of zero-field spin photocurrent from magnetoelectric photocurrent}

\begin{figure}[tbph]
\centering \includegraphics[width=0.5\textwidth]{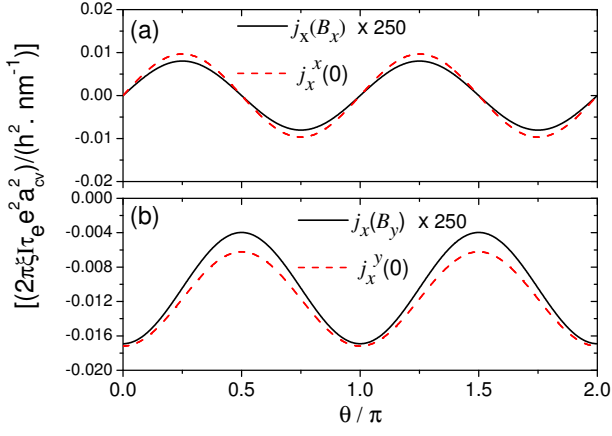}
\caption{(Color online) Numerical comparisons between the magnetoelectric photocurrents and zero-field pure spin photocurrents. (a) $j_x(B_y)\times 250$ at $B_y=1$ T and $j^x_y(0)$. (b) $j_x(B_x)\times 250$ at $B_x=1$ T and $j^x_x(0)$. The parameters are given in Sec. \ref{sec:quantitative}. For a direct comparison, we neglect the $-e$ and $\hbar/2$ in front of the current density and spin current density formulas. The above values of currents are in units of $(2\pi \xi I \tau_e e^2 a_{cv}^2)/(h^2\cdot \mathrm{nm})$, and have a dimension of second$^{-1}$meter$^{-1}$.}
\label{fig:compare}
\end{figure}

Because both the magnetoelectric photocurrents and the zero-field pure spin photocurrents originate from the same photoexcited carrier density, this allows us to find a relation between them. With the help of Eqs. (\ref{jmubnuapprox}) and (\ref{jxxresult}), the ratio of the zero-field longitudinal pure spin photocurrent to the electric photocurrent induced by the parallel magnetic field turns out to be
\begin{eqnarray}\label{jxBxjxx0ratio}
\left|\frac{j^x_x(0)}{j_x(B_x)}\right|&\simeq& \left|\frac{\sum_{c,v}c\rho^{\sin}_{cv}  \frac{\hbar^2 k^2_{cv}}{2m^*}}{\eta h_x \sum_{c,v} c\rho^{\sin}_{cv} }\right|,
\end{eqnarray}
and the ratio of the transverse spin photocurrent to the perpendicular magnetic field induced electric photocurrent is given by \begin{eqnarray}\label{jxByjxy0ratio}
\left|\frac{j_{x}^{y}(0)}{j_x(B_y)}\right| &\simeq &
\frac{\left|\sum_{c,v} c\rho^0_{cv}\frac{\hbar^2k^2_{cv}}{2m^*} + \alpha \rho^0_{cv}k_{cv}+\frac{1}{2}\cos2\theta  c\rho^{\cos}_{cv}\frac{\hbar^2k^2_{cv}}{2m^*} \right|}{ \left|\eta h_y \sum_{c,v} \left[ c\rho^0_{cv}-\frac{1}{2}\cos2\theta c\rho^{\cos}_{cv}\right]\right|}\nonumber\\
&\approx &
\frac{\left|\sum_{c,v}\left(c\rho^0_{cv}\frac{\hbar^2k^2_{cv}}{2m^*} \right)\right|}{ \left|\eta h_y \sum_{c,v}  c\rho^0_{cv}\right|}.
\end{eqnarray}
These relations have clear physical meaning. Note that $\sum_{c=\pm}c\rho^{0,\cos,\sin}_{cv}=\rho^{0,\cos,\sin}_{+v}-\rho^{0,\cos,\sin}_{-v}$. This difference of the photoexcited carrier density between the $+$ and $-$ conduction bands, as already shown in Fig. \ref{fig:rho0}, is due to the spin-orbit coupling. Therefore, the denominators in Eqs. (\ref{jxBxjxx0ratio}) and (\ref{jxByjxy0ratio}) mean that both the magnetic field and the spin-orbit coupling are necessary ingredients of the magnetoelectric photocurrent, while the numerators indicate that the pure spin currents are proportional to the spin-orbit coupling and the kinetic energy of the photoexcited carriers (if we can view literally $\hbar^2k^2_{cv}/2m^*$ as kinetic energy).

At this moment, we make a bold approximation by canceling the effect of the spin-orbit coupling from both the denominators and numerators in Eqs. (\ref{jxBxjxx0ratio}) and (\ref{jxByjxy0ratio}), and literally say that the ratio of the zero-field pure spin photocurrent to the magnetoelectric photocurrent is about ``\emph{kinetic energy over Zeeman energy}".
This relation, though rather coarse, can help us to make a quick order-of-magnitude estimate of the undetectable pure spin photocurrent from the measured magnetoelectric photocurrent.\cite{Dai2010.prl.104.246601} The average kinetic energy of the photoexcited carriers is higher than the equilibrium Fermi energy measured from the bottom of the conduction bands, and thus is more than $E_{\mathrm{F}}=0.01$ eV for our numerical calculations. The Zeeman energy induced by 1 tesla of magnetic field is about $10^{-4}$ eV for the Land\'{e} \emph{g}-factor of $g_e=-4$.\cite{Smith1987.prb.35.7729,Nitta1997.prl.78.1335,Winkler} Therefore, the rough estimate implies that the spin photocurrent is about two orders larger than the magnetoelectric photocurrent at 1 tesla. To test the reliability of this quick estimate, we numerically compare the magnetoelectric photocurrent and zero-field pure spin photocurrent in Fig. \ref{fig:compare}.
The spin photocurrents are about $250$ times larger than the magnetoelectric photocurrent at 1 tesla of magnetic field. Despite its roughness, this quick estimate gives a reasonable result. We expect that this quick estimate can serve as a reference for more sophisticated non-destructive approaches, such as the Faraday rotation\cite{Wang2008.prl.100.086603} or the second-order nonlinear optical effects.\cite{Wang2010.prl.104.256601,Werake2010}

\section{\label{sec:conclusion}Conclusions}
In this work, we present a theoretical description of the recent experiment\cite{Dai2010.prl.104.246601} on the optical injection of spin-polarized carriers via direct interband optical excitations into a semiconductor quantum well under the normal incidence of linearly polarized or unpolarized lights. In that experiment, the injection produces pure spin photocurrents accompanying no electric current at zero magnetic field due to spin-orbit coupling that respects time-reversal symmetry. An in-plane magnetic field can break time-reversal symmetry, and extracts a magnetoelectric photocurrent. The magnetoelectric photocurrent is characterized by its dependence on the magnetic fields and the polarization of the light.

With the help of the density matrix formalism, we calculate the photoexcited carrier density, current density, and spin current density.
The photoexcited carrier density in $\mathbf{k}$ space shows an anisotropic dependence on both the wave vector angle $\varphi$ and the polarization angle $\theta$ of the linearly polarized light as given in Eq. (\ref{eq:rho0}). Since the velocities of carriers can also be expressed as functions of $\varphi$ [see Eq. (\ref{velocitytaylor})],
we can show that the current density can be simplified as an angle integral over $\varphi$ of the product of the density and velocity of photoexcited carriers. The angle integral then produces all the magnetic field and polarization angle dependences of the magnetoelectric photocurrent as reported in Ref. \onlinecite{Dai2010.prl.104.246601}, in particular, the magnetoelectric photocurrent induced by the parallel magnetic field.

We show that the present simplified model with Rashba and Luttinger Hamiltonians is able to reproduce the current formula for the magnetoelectric photocurrent. However, discrepancies still exist between theory and experiment for the relative magnitudes and signs among the parameters  in the current formula. Since the simplified model is a natural choice considering the symmetry of the quantum well that was investigated in the experiment, the discrepancies indicate that further investigations, with more detailed considerations on the band structure of the sample, are needed to identify the origin of the discrepancies. For example, more energy bands may be required in the calculation, the graded doping in the quantum well may need a self-consistent calculation of the potential and carrier density.

We further show that the origin of the previously predicted pure spin photocurrents\cite{Li2006.apl.88.162105,Zhou2007.prb.75.045339} can be well illustrated from the same photoexcited carrier density. We propose that the ratio of the zero-field pure spin photocurrent to the magnetoelectric photocurrent can be approximated as ``\emph{kinetic energy over Zeeman energy}". With this relation, the underlying pure spin photocurrent can be quickly estimated from the observed magnetoelectric photocurrent, and provides a reference for other approaches to nondestructive measurement of the pure spin photocurrent.

\section{Acknowledgments}

We thank Xiaodong Cui, Junfeng Dai, Chun-Lei Yang, Wei-Qiang Chen, Jing Wang, Ren-Bao Liu, and Bang-fen Zhu for helpful discussions. This work was supported by the Research Grant Council of Hong Kong under Grant Nos. HKU7041/07P, and HKU 10/CRF/08. Z.B. was supported by National Natural Science Foundation of China (Grant No. 10974046) and Hubei Provincial Natural Science Foundation of China (Grant No. 2009CDB360).

\appendix

\section{\label{sec:C2v}$C_{\mathrm{2v}}$ group and anisotropic Rashba model}

\begin{figure}[htbp]
\centering
\includegraphics[width=0.45\textwidth]{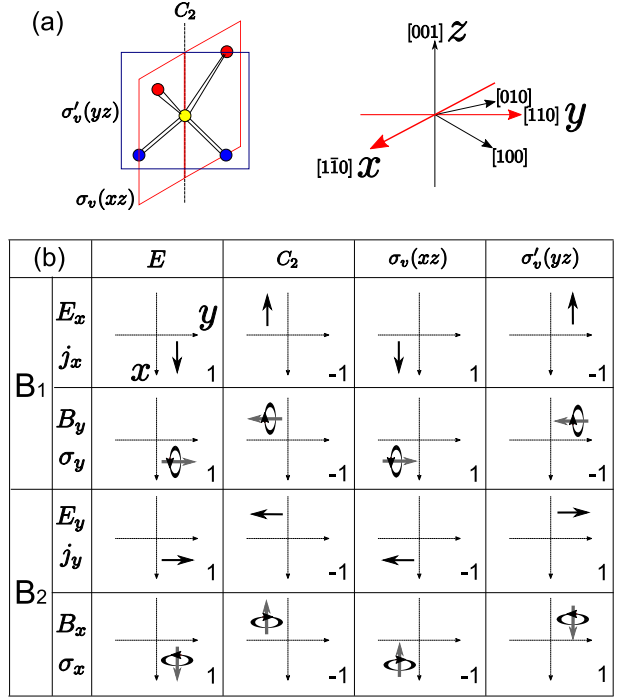}
\caption{(Color online) (a) The building block of an inversion-asymmetric zinc blende structure grown along [001] crystallographic direction. When $x$ and $y$ axes are defined along [1\={1}0] and [110] crystallographic directions, the $xz$ and $yz$ planes coincide with the mirror reflection planes of the $C_{\mathrm{2v}}$ point group. (b) How polar vectors $j_x$, $j_y$, $E_x$, $E_y$ and axial vector $B_x$, $B_y$, $\sigma_x$, $\sigma_y$ transform under the four symmetry operations of the $C_{\mathrm{2v}}$ point group. $E$: identical; $C_2$: two-fold rotation about $z$ axis; $\sigma_{v}$ ($\sigma_{v}'$): mirror reflection with respect to $xz$ ($yz$) plane. Whether the vector changes sign under the symmetry operation is indicated on the lower right corner of each panel by ``1" or ``-1", which are actually the characters of the representations B$_1$ and B$_2$ of the $C_{\mathrm{2v}}$ group.\cite{Dresselhaus2008.Book.GroupTheory}}\label{fig:C2v}
\end{figure}

It is well known that the heterostructures of inversion-asymmetric zinc blende materials grown along the [001] direction have the $C_{\mathrm{2v}}$ point group symmetry. The basic building block of these structures is shown in Fig. \ref{fig:C2v}(a). It has four symmetry operations. When the $x$ and $y$ axes are defined as [1\={1}0] and [110] crystallographic directions, respectively, the $xz$ and $yz$ planes coincide with the mirror reflection planes of the $C_{\mathrm{2v}}$ group. As the basis functions, the polar vectors (such as
velocity, current, electric field) along the $x$ axis and the axial vectors (such as spin and magnetic field) along the $y$ axis transform according to the irreducible representation B$_{1}$ of the $C_{\mathrm{2v}}$ group, and the polar vectors along the $y$ and the axial vectors along the $x$ directions according to the irreducible representation B$_{2}$.\cite{Dresselhaus2008.Book.GroupTheory} The physical picture of ``vectors transforming according to irreducible representations" is schematically illustrated in Fig. \ref{fig:C2v}(b).

\begin{table}[tbph]
\caption{Two examples of how to determine whether an element of the pseudo tensor $\chi^{\alpha\beta\gamma\delta}$ is nonzero. $1$ and $-1$ correspond to ``remaining unchanged" and ``changing sign", respectively, upon applying the symmetry operations of the $C_{\mathrm{2v}}$ group to the vectors. (a) For $\chi^{xxxx}$, which is zero because $j_{x}$ and $B_xE_xE_x$ are different for $\sigma_v$ and $\sigma_{v}'$. (b) For $\chi^{xxxy}$, which is nonzero because $j_{x}$ and $B_xE_xE_y$ are the same for all symmetry operations.}
\label{tab:c2v}
\begin{tabular}{c c c c c c}
  \hline
  \hline
   (a) & $j_x$ & $B_x$ & $E_x$ & $E_x$ & $B_xE_xE_x$ \\
     \hline
 $C_2$    & -1 & -1 & -1 & -1 & -1 \\
 $\sigma_v$    & 1 & -1 & 1 & 1 & -1 \\
 $\sigma_v'$    & -1 & 1 & -1 & -1 & 1 \\
  \hline
  \hline
   (b)  & $j_x$ & $B_x$ & $E_x$ & $E_y$ & $B_xE_xE_y$ \\
   \hline
  $C_2$  & -1 & -1 & -1 & -1 & -1 \\
 $\sigma_v$  & 1 & -1 & 1 & -1 & 1 \\
 $\sigma_v'$  & -1 & 1 & -1 & 1 & -1 \\
  \hline
  \hline
\end{tabular}
\end{table}

The basic observation to the experiment data indicates that the current is linearly proportional to the magnetic field, and the $2\theta$ function dependence usually implies a second-order nonlinear optics. Phenomenologically, the current density $j_{\alpha}$ can be generally written as\cite{Belkov2005.jpcm.17.3405}
\begin{eqnarray}\label{jphenomenalogical}
j_{\alpha} &=& \chi^{\alpha\beta\gamma\delta}B_{\beta}E_{\gamma}E_{\beta},
\end{eqnarray}
where $\alpha$, $\beta$, $\gamma$, and $\delta$ stand for Cartesian coordinates. $B_{\beta}$ and $E_{\gamma}$ are the components of the magnetic field and polarization electric field vector. The nonzero terms of Eq. (\ref{jphenomenalogical}) require that the vectors on both sides transform in the same manner for all the symmetry operations of the $C_{\mathrm{2v}}$ group. Two examples are illustrated in Table \ref{tab:c2v}. In the language of the irreducible representation of group theory, the table \ref{tab:c2v} can be written as
\begin{eqnarray}
B_1 &\neq & B_2\otimes B_1 \otimes B_1 , \nonumber\\
B_1 & = & B_2\otimes B_1 \otimes B_2  .
\end{eqnarray}
Similarly, all the nonzero terms can be found and summarized as Eq. (\ref{eq:Jsymmetry}).

In addition, both $\sigma_x$ and $k_y$ transform according to B$_1$, while B$_1$ $\otimes$ B$_1$ yields the identity representation of the $C_{\mathrm{2v}}$ group, so
$\sigma_xk_y$ is an invariant for structures with the $C_{\mathrm{2v}}$ symmetry. Similarly, $\sigma_y k_x$ is also an invariant. On the contrary, $\sigma_xk_x$ and $\sigma_yk_y$ are not invariants. As a result, the spin-orbit coupling up to the linear order in $\mathbf{k}$ can be generally described by the form
\begin{eqnarray}
H_{\mathrm{SOC}} = \lambda_y \sigma_x k_y - \lambda_x\sigma_y k_x .
\end{eqnarray}
where $\lambda_y$ and $\lambda_x$ are spin-orbit coupling coefficients along different directions.
In the coordinate system where $x'||[100]$ and $y'||[010]$, the spin-orbit coupling in the conduction bands of a sample with the $C_{\mathrm{2v}}$ symmetry can be generally written as\cite{Ganichev2004.prl.92.256601,Giglberger2007.prb.75.035327}
\begin{eqnarray}
H_{\mathrm{SOC}}' &=& \alpha(\sigma_{x'} k_{y'} -\sigma_{y'} k_{x'}) + \beta (\sigma_{x'} k_{x'} -\sigma_{y'} k_{y'})
\end{eqnarray}
referred to as Rashba ($\alpha$) and Dresselhaus ($\beta$) terms, respectively. While in this work the coordinate system is $x||[1\bar{1}0]$ and $y||[110]$, the form of the spin-orbit coupling can be obtained by rotating the above $H_{\mathrm{SOC}}'$ by 45$^{\circ}$
\begin{eqnarray}\label{dresselhaus}
H_{\mathrm{SOC}} &=& (\alpha+\beta)\sigma_xk_y -(\alpha-\beta)\sigma_yk_x.
\end{eqnarray}
The extra $\beta$ means that although the $C_{\mathrm{2v}}$ symmetry allows only $\sigma_xk_y$- or $\sigma_yk_x$-type spin-orbit coupling when $x||[1\bar{1}0]$ and $y||[110]$, the Dresselhaus term when $x'||[100]$ and $y'||[010]$ can lead to an anisotropy of the spin-orbit coupling. Usually, $\beta$ is smaller than $\alpha$.\cite{Ganichev2004.prl.92.256601,Giglberger2007.prb.75.035327} The anisotropy of the spin-orbit coupling may explain why $c_x\neq c_y$ in the experiment.\cite{Dai2010.prl.104.246601}
Note that this result applies only for when $x$ and $y$ directions are referred to the [1\={1}0] and [110] crystallographic directions.\cite{Ganichev2004.prl.92.256601,Giglberger2007.prb.75.035327}

\section{\label{sec:Numerical} Numerical scheme}

For each pair of conduction ($c$) and valence ($v$) bands, we denote all the quantum states by the discrete values of the wave vector angle $\varphi\rightarrow \varphi_i=i2\pi/N$, with $i=0,1,...,N-1$. At each $\varphi_i$, the energy conservation law $\omega=\omega_{cv}(k,\varphi_i)$ is numerically solved. The root of $k$ is the wave vectors $k_{cv}(\varphi_i)$ on the constant energy contours. The physical quantities such as $v^{x}_{cv,\mathbf{k}}$ and $\rho^{0,\cos,\sin}_{cv,\mathbf{k}}$ then are calculated by giving them $\mathbf{k}=k_{cv}(\varphi_i)(\cos\varphi_i,\sin\varphi_i)$, and denoted as $v^{x}_{cv,\mathbf{k}}(\varphi_i)$ and $\rho^{0,\cos,\sin}_{cv,\mathbf{k}}(\varphi_i)$.
As an example, we show how to numerically calculate the $\theta$-independent term of $j_x$.
According to Eqs. (\ref{jxpolar}) and (\ref{rho0deltak}), it means that we calculate
\begin{eqnarray}\label{j0sum}
j_{x}^0
 &=& -\frac{\xi I  \tau_e  e^3a_{\mathrm{cv}}^2\sqrt{\mu_0/\varepsilon_0} }{2\pi\hbar^2}\sum_{c,v}
\sum_{i=0}^{N-1}
 \Delta k_{cv}(\varphi_i)
v^x_{cv,\mathbf{k}}(\varphi_i)\nonumber\\
&\times&\{1-f_{c}[k_{cv}(\varphi_i)]\}\left.\left(\frac{|x_{cv}|^2+|y_{cv}|^2}{a_{\mathrm{cv}}^2}\right)\right|_{k_{cv}(\varphi_i)}G [k_{cv}(\varphi_i)],\nonumber\\
\end{eqnarray}
where $\Delta k_{cv}(\varphi_i)\equiv \frac{2\pi}{N} k_{cv}(\varphi_i)=|\mathbf{k}_{cv}(\varphi_i)-\mathbf{k}_{cv}(\varphi_{i-1})|$, and $\sum_{i=0}^{N-1}
 \Delta k_{cv}(\varphi_i)$ gives the circumference
of the constant energy contour. In the summation, we are using the units $[G]\sim $ second/meter, $[v^x]\sim $ meter/second,
$[\Delta k_{cv}]\sim $ nm$^{-1}$, so the units of the current density in front of the dimensionless summation of Eq. (\ref{j0sum}) are
\begin{eqnarray}\label{j0units}
[j_{x}^0]
 &\sim&  \frac{  \xi I \tau_e  e^3  a_{\mathrm{cv}}^2 \sqrt{\mu_0/\varepsilon_0}}{2\pi\hbar^2 \mathrm{nm}}
  \approx  0.09 \ \mathrm{A/m}.
\end{eqnarray}

\end{CJK*}

\end{document}